\documentclass[english]{article}
\usepackage[T1]{fontenc}
\usepackage[latin9]{inputenc}
\usepackage{rotating}
\usepackage{float}
\usepackage{multirow}
\usepackage{amsmath}
\usepackage{amssymb}
\usepackage{graphicx}
\usepackage{setspace}

\makeatletter

\providecommand{\tabularnewline}{\\}
\floatstyle{ruled}
\newfloat{algorithm}{tbp}{loa}
\providecommand{\algorithmname}{Algorithm}
\floatname{algorithm}{\protect\algorithmname}

\makeatother

\usepackage{babel}
\begin{document}
\title{A Low-Memory Time-Efficient Implementation of Outermorphisms for Higher-Dimensional
Geometric Algebras}
\author{Ahmad Hosny Eid}
\maketitle
\begin{abstract}
From the beginning of David Hestenes rediscovery of geometric algebra
in the 1960s, outermorphisms have been a cornerstone in the mathematical
development of GA. Many important mathematical formulations in GA
can be expressed as outermorphisms such as versor products, linear
projection operators, and mapping between related coordinate frames.
Over the last two decades, GA-based mathematical models and software
implementations have been developed in many fields of science and
engineering. As such, efficient implementations of outermorphisms
are of significant importance within this context. This work attempts
to shed some light on the problem of optimizing software implementations
of outermorphisms for practical prototyping applications using geometric
algebra. The approach we propose here for implementing outermorphisms
requires orders of magnitude less memory compared to other common
approaches, while being comparable in time performance, especially
for high-dimensional geometric algebras.
\end{abstract}
Geometric Algebra, Outermorphism, Software Implementation

\section{Background}

In geometric algebra, the outermorphism $\overline{\mathbf{T}}$ of
a linear map $\mathbf{T}$ between two vector spaces is an extension
of the linear map that acts on arbitrary multivectors of geometric
algebras constructed on the two vector spaces \cite{Hestenes1987}.
From the beginning of the rediscovery of GA in the 1960s, outermorphisms
have been a cornerstone in the development of GA. Many important mathematical
formulations and operators in GA can be expressed as outermorphisms.
Such formulations include the versor product, rotors, and linear projection
operators, among many others \cite{Dorst2009,Perwass2008}. In addition,
outermorphisms provide for a suitable approach for performing common
products on multivectors within non-orthogonal coordinate frames \cite{Eid2018}.

Over the last two decade, GA have matured to enter many practical
applications in science and engineering \cite{Hitzer2013,Hildenbrand2015,Bayro-Corrochano2018,Joot2019,Wang2019}.
As such, computational aspects of GA are becoming more important for
investigating and prototyping mathematical and computational models
based on GA mathematics. As a core part of GA, efficient implementations
of outermorphisms are of significant importance in this context. Unfortunately,
most optimization efforts targeting efficient GA implementations mainly
focus on optimizing core products on multivectors, such as the geometric,
outer, and inner products. Although the optimization of products is
important, the focus on outermorphism-related computations is equally
important in many practical cases, as a single outermorphism can replace
several product operations on multivectors.

This work focuses on the problem of optimizing software implementations
of outermorphisms for practical prototyping applications using geometric
algebra. The approach we propose here for implementing outermorphisms
requires orders of magnitude lower memory compared to common approaches,
while being comparable in time performance. The main benefit of the
approach we propose in this work appears in higher-dimensional GAs
with dimensions larger than 12, where common approaches become infeasible
due to large memory requirements.

This section gives the necessary background to formulate the proposed
approach including the definition of GA Coordinate Frames (GACFs),
the use of binary trees to represent multivectors, and the definition
of outermorphisms on GACFs. Section 2 explains relevant algorithmic
and implementational details of the proposed approach for efficiently
mapping multivectors using outermorphisms. Section 3 illustrates the
usefulness of the proposed approach using several experiments and
a brief discussion of the results. Finally, section 4 provides conclusions
to this work.

\subsection{Geometric Algebra Coordinate Frames}

In this work, a Geometric Algebra Coordinate Frame (GACF) \cite{Eid2018}
$\mathcal{F}\left(\boldsymbol{F}_{1}^{n},\mathbf{A}_{\mathcal{F}}\right)$
is the mathematical structure used to define computations on a geometric
algebra $\mathcal{G}^{p,q,r}$ in terms of basic scalar coordinates
commonly used to implement computations on a computer. The GACF framework
is a reformulation and extension of the computational GA framework
provided in \cite{Dorst2009} to uniformly work with orthogonal and
non-orthogonal GA coordinate frames alike in practice. A GACF is completely
defined using two components:
\begin{enumerate}
\item An ordered set of $n$ basis vectors $\boldsymbol{F}_{1}^{n}=\left\langle f_{0},f_{1},\cdots,f_{n-1}\right\rangle $
defining the dimensions of the GACF's base vector space.
\item A symmetric real bilinear form $\mathbf{B}:\boldsymbol{F}_{1}^{n}\times\boldsymbol{F}_{1}^{n}\rightarrow\mathbb{R},\,\mathbf{B}\left(f_{i},f_{j}\right)=\mathbf{B}\left(f_{j},f_{i}\right)=f_{i}\cdot f_{j}$
determining the inner product of basis vectors and given by a symmetric
$n\times n$ bilinear form matrix $\mathbf{A}_{\mathcal{F}}=\left[f_{i}\cdot f_{j}\right]$
called the Inner Product Matrix (IPM) of the GACF.
\end{enumerate}
A GACF can be of two types: orthogonal or non-orthogonal. The IPM
of an orthogonal GACF is diagonal ($f_{i}\cdot f_{i}=d_{i}$, $f_{i}\cdot f_{j}=0\,\forall i\neq j$)
while the IPM of a non-orthogonal GACF is non-diagonal ($f_{i}\cdot f_{j}=f_{j}\cdot f_{i}=b_{ij}\,\exists i\neq j:b_{ij}\neq0$).
A Euclidean GACF is orthogonal with all $d_{i}=1$.

We construct three additional components to serve GA computations
within the GACF:
\begin{enumerate}
\item The ordered set of $2^{n}$ basis blades of all grades $\boldsymbol{F}^{n}=\left\langle F_{0},F_{1},\cdots,F_{2^{n}-1}\right\rangle $.
This set is automatically determined by the set of basis vectors $\boldsymbol{F}_{1}^{n}$.
This component is independent of the metric represented by $\mathbf{A}_{\mathcal{F}}$
as they are created using the metric-independent outer product of
basis vectors in $\boldsymbol{F}_{1}^{n}$:
\begin{eqnarray}
F_{i} & = & \prod_{\wedge}\left(\boldsymbol{F}_{1}^{n},i\right)\\
 & = & \begin{cases}
1 & ,i=0\\
f_{m} & ,i=2^{m},m\in\{0,1,\cdots,n-1\}\\
f_{i_{1}}\wedge f_{i_{2}}\wedge\cdots\wedge f_{i_{r}} & ,\begin{array}{c}
i=2^{i_{1}}+2^{i_{2}}+\cdots+2^{i_{r}},\\
i_{i}<i_{2}<\cdots<i_{r}
\end{array}
\end{cases}\nonumber 
\end{eqnarray}
\item The geometric product; a bilinear operator on multivectors $G_{\mathcal{F}}:\boldsymbol{F}^{n}\times\boldsymbol{F}^{n}\rightarrow\mathcal{G}^{p,q,r}$
defined using the geometric product of pairs of basis blades, which
are generally multivectors, $G_{\mathcal{F}}(F_{i},F_{j})=F_{i}F_{j}=\sum_{k=0}^{2^{n}-1}g_{k}F_{k},\,g_{k}\in\mathbb{R}$.
This bilinear operator is automatically determined by the set of basis
vectors $\boldsymbol{F}_{1}^{n}$ and the bilinear form $\mathbf{B}$.
\item If the bilinear form is not orthogonal, a base orthogonal GACF $\mathcal{E}\left(\boldsymbol{E}_{1}^{n},\mathbf{A}_{\mathcal{E}}\right)$
of the same dimension is needed, in addition to an orthogonal Change-of-Basis
Matrix (CBM) $\mathbf{C}$. The orthogonal CBM is used to express
basis vectors of $\mathcal{F}$ as linear combinations of basis vectors
of $\mathcal{E}$, and defines a Change of Basis Automorphism (CBA)
$\overline{\mathbf{C}}$ that can invariantly transform linear operations
on multivectors between $\mathcal{E}$ and $\mathcal{F}$. We can
either define $\mathbf{C}$ implicitly from the orthonormal eigen
vectors of $\mathbf{A}_{\mathcal{F}}$, or the user can directly supply
$\mathcal{E}\left(\boldsymbol{E}_{1}^{n},\mathbf{A}_{\mathcal{E}}\right)$
and $\mathbf{C}$ to define the IPM of $\mathcal{F}$. The details
of this component are described in \cite{Eid2018}.
\end{enumerate}
Using these five components any multivector $X=\sum_{i=0}^{2^{n}-1}x_{i}F_{i},\,x_{i}\in\mathbb{R}$
can be represented by a column vector of real coefficients $\left[x_{i}\right]_{\mathcal{F}}$.
All common operations on multivectors are easily encoded using this
framework for both orthogonal and non-orthoogonal GACFs. Such operations
include common bilinear products, outermorphisms, and versor transformations,
as detailed in \cite{Eid2018}.

\subsection{Binary Tree Representation of Multivectors}

\begin{figure}
\noindent \begin{centering}
\includegraphics[width=7.5cm]{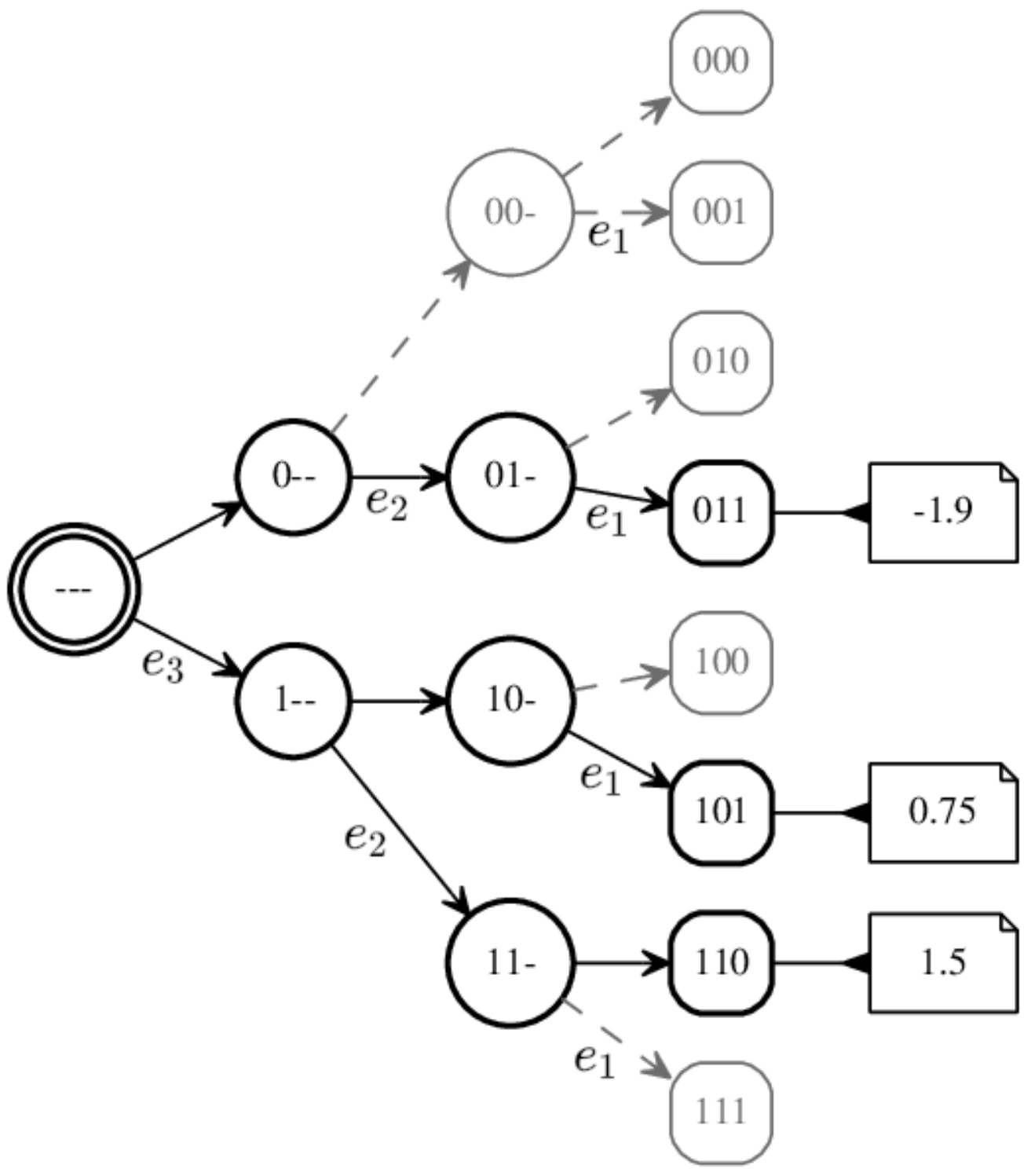}
\par\end{centering}
\caption{Example for BTR of the multivector $-1.9e_{12}+0.75e_{13}+1.5e_{23}$
defined on a GACF with basis vectors $\left\langle e_{1},e_{2},e_{3}\right\rangle $.
Tree nodes and edges stored in memory are denoted using black solid
lines. Grey nodes and dashed edges are only shown for illustration
and not stored in memory for this multivector. Node IDs and tree depths
are shown inside the nodes. For example, the internal node denoted
by '$01-$' has an ID of $(010)_{2}=2$ and a tree depth of $1$,
while the internal node '$1--$' has an ID of $(100)_{2}=4$ and a
tree depth of $2$.}
\label{fig:tree-multivector}
\end{figure}

We begin from an arbitrary GACF $\mathcal{E}\left(\boldsymbol{E}_{1}^{n},\mathbf{A}_{\mathcal{E}}\right)$
defined on the geometric algebra $\mathcal{G}^{p,q,r}$ with $n=p+q+r$
basis vectors $\boldsymbol{E}_{1}^{n}=\left\langle e_{0},e_{1},\cdots,e_{n-1}\right\rangle $.
A multivector $X=\sum_{i=0}^{2^{n}-1}x_{i}E_{i}$ is a linear combination
of basis blades $E_{i}$ in $\mathcal{E}$. Here we use a Binary Tree
Representation (BTR) of $X$ very similar to, and inspired by, the
one first proposed in \cite{Fuchs2014} and later developed in \cite{Breuils2017,Breuils2017a,breuils:tel-02085820}.
As seen in Figure \ref{fig:tree-multivector}, the main difference
lies in the ordering of basis vectors in tree levels. In \cite{Fuchs2014},
basis vectors are introduced in the tree starting from root to leafs
in the order $e_{0},e_{1},\cdots,e_{n-1}$. In this work, however,
the order is reversed $e_{n-1},e_{n-2},\cdots,e_{0}$. This reversal
of basis vectors order is significant for it enables the possibility
of efficiently embedding smaller trees into larger ones, explained
shortly, thus reusing the same tree for several related multivectors.
One other difference with the approach of \cite{Fuchs2014} is that
some additional information are stored in tree nodes to speed-up computations
on multivectors as described next.

To understand tree reuse this structure provides, assume as an example
we have a Euclidean position vector $A=xe_{1}+ye_{2}+ze_{3}$ that
we wish to represent as a conformal multivector $A_{C}=n_{o}+A+\frac{1}{2}A^{2}n_{\infty}$.
Because the Euclidean GA multivector $A$ is actually part of its
conformal representation $A_{C}$, we can directly utilize the in-memory
BTR of $A$ without any changes as a sub-tree of the BTR of $A_{C}$.
For large multivectors this organization would significantly save
memory, and enables caching multivector computations for later use.

Nodes in the BTR are of two kinds: internal nodes and leaf nodes.
Actual multivector data, basis blade IDs $i$ and scalar coefficients
$x_{i}$, reside in leaf nodes. Internal nodes are only used to efficiently
guide calculations inside computational procedures on multivectors.
A leaf node $N_{X}$ essentially holds 2 pieces of information: an
integer-valued basis blade ID $\mathtt{ID}\left(N_{X}\right)$, and
the associated scalar coefficient $\mathtt{ScalarValue}\left(N_{X}\right)$;
a floating-point number. An internal node $N_{X}$ holds 4 pieces
of information: its integer-valued tree depth $\mathtt{TreeDepth}\left(N_{X}\right)$,
its node ID $\mathtt{ID}\left(N_{X}\right)$, and two memory references
to child nodes; either of them can be null, but not both. Each internal
node can have a 0-child $\mathtt{Child_{0}}\left(N_{X}\right)$, a
1-child $\mathtt{Child_{1}}\left(N_{X}\right)$, or both. Tree depth
of an internal node is the number of tree levels under the node. The
root internal node in Figure \ref{fig:tree-multivector}, for example,
has a tree depth of 3, which is the same as the number of basis vectors
of the GACF. In any such tree, the tree depth of internal nodes in
the level just before the leaf nodes is always 1. The internal node's
ID is used to compute its two child nodes' IDs. The ID of a 0-child
is equal to the ID of its parent internal node. The ID of a 1-child
is equal to the ID of its parent plus $2^{d-1}$, where $d$ is the
tree depth of the parent internal node. Although it is always possible
to compute all node IDs on-the-fly during computations, we prefer
to store them inside tree nodes to save some processing time.

\subsection{Outermorphisms on GACFs}

In this work, we will consider outermorphisms between GACFs, not GAs,
with no loss of generality. An outermorphism $\overline{\mathbf{T}}:\mathcal{E}\rightarrow\mathcal{F}$
is a linear map on multivectors defined between two GACFs $\mathcal{E}\left(\boldsymbol{E}_{1}^{n},\mathbf{A}_{\mathcal{E}}\right)$,
$\mathcal{F}\left(\boldsymbol{F}_{1}^{m},\mathbf{A}_{\mathcal{F}}\right)$
where $\overline{\mathbf{T}}\left[A\wedge B\right]=\overline{\mathbf{T}}\left[A\right]\wedge\overline{\mathbf{T}}\left[B\right]$,
$\overline{\mathbf{T}}\left[A+B\right]=\overline{\mathbf{T}}\left[A\right]+\overline{\mathbf{T}}\left[B\right]$,
and $\overline{\mathbf{T}}\left[\alpha A\right]=\alpha\overline{\mathbf{T}}\left[A\right]$
for all multivectors $A,B$ and scalars $\alpha$. The two GACFs could
represent the same GA, thus $m=n$, or two different GAs if needed.
If both GACFs represent the same GA, either with similar or different
basis blades, the outermorphism $\overline{\mathbf{T}}$ is a linear
operator on the GA.

Because any GA is essentially a linear space with additional structure,
we can fully define any linear map $\mathbf{L}$ on multivectors if
we know the effect of the map on the basis blades of the domain GACF
$L_{i}=\mathbf{L}\left[E_{i}\right],i=0,1,\ldots,2^{n}-1$, where
$L_{i}=\sum_{j=0}^{2^{m}-1}l_{i,j}F_{j}$ are multivectors defined
as linear combinations of basis blades $\left\langle F_{0},F_{1},\cdots,F_{2^{m}-1}\right\rangle $
in $\mathcal{F}$. This is easily extended by linearity to any multivector
$X=\sum_{i=0}^{2^{n}-1}x_{i}E_{i}$ to compute its map $\mathbf{L}\left[X\right]=\sum_{i=0}^{2^{n}-1}x_{i}L_{i}$.

For an outermorphism $\overline{\mathbf{T}}$ we can fully define
the map just by knowing its effect on the domain GACF basis vectors
$\boldsymbol{t}_{i}=\overline{\mathbf{T}}\left[e_{i}\right],i=0,1,\ldots,n-1$,
where $\boldsymbol{t}_{i}=\sum_{j=0}^{m-1}t_{i,j}f_{j}$ are vectors
exclusively defined as linear combinations of basis vectors in $\mathcal{F}$.
To find the outermorphism $\overline{\mathbf{T}}\left[E_{i}\right]$
of an arbitrary basis blade $E_{i}=e_{i_{0}}\wedge e_{i_{1}}\wedge\cdots\wedge e_{i_{k}}$
of grade $k$, we can simply use: 
\begin{eqnarray}
T_{i} & = & \overline{\mathbf{T}}\left[E_{i}\right]\nonumber \\
 & = & \overline{\mathbf{T}}\left[e_{i_{0}}\wedge e_{i_{1}}\wedge\cdots\wedge e_{i_{k}}\right]\nonumber \\
 & = & \overline{\mathbf{T}}\left[e_{i_{0}}\right]\wedge\overline{\mathbf{T}}\left[e_{i_{1}}\right]\wedge\cdots\wedge\overline{\mathbf{T}}\left[e_{i_{k}}\right]\nonumber \\
 & = & \boldsymbol{t}_{i_{0}}\wedge\boldsymbol{t}_{i_{1}}\wedge\cdots\wedge\boldsymbol{t}_{i_{k}}\label{eq:kvectors}
\end{eqnarray}

We can then use linear extension $\overline{\mathbf{T}}\left[X\right]=\sum_{i=0}^{2^{n}-1}x_{i}T_{i}$
to map arbitrary multivectors as before, while noting that $T_{0}=\overline{\mathbf{T}}\left[E_{0}\right]=\overline{\mathbf{T}}\left[1\right]=1$
for all outermorphisms. We note from relation \ref{eq:kvectors} that
the outermorphism of a basis blade of grade $k$ in $\mathcal{E}$
is always a $k$-blade (a $k$-vector) in $\mathcal{F}$, which might
be zero in some cases.

For each outermorphism $\overline{\mathbf{T}}$, typical software
implementations pre-compute and store its mapped $k$-vectors $T_{i}=\overline{\mathbf{T}}\left[E_{i}\right]$
in computer memory using various forms, including a full or sparse
matrix of size $2^{m}\times2^{n}$. A related and common approach
is to store $T_{i}$ inside a set of $m+1$ square matrices $\left\{ M_{0},M_{1},\ldots,M_{m}\right\} $
where the size of $M_{k}$ is $\left(\begin{array}{c}
n\\
k
\end{array}\right)\times\left(\begin{array}{c}
n\\
k
\end{array}\right)$ as described, for example, in \cite{Dorst2009} and \cite{Eid2018}.
We will designate this class of methods as Cached Basis-Mapping Methods
(CBMMs). For high-dimensional GAs CBMMs are generally infeasible because
of memory size constraints. For an arbitrary outermorphism operator
on a 15-dimensional GA, for example, we would need at least $8\sum_{k=0}^{15}\left(\begin{array}{c}
15\\
k
\end{array}\right)^{2}=1,240,940,160\simeq1.15$ GBytes in memory, when using double precision floating point numbers.
The situation is much worse for higher-dimensional GAs, especially
when several outermorphisms are needed for computations. In addition,
typical multivectors are highly sparse in most practical applications,
especially for higher-dimensional GAs. The use of matrices to represent
outermorphisms doesn't exploit such sparsity resulting in unnecessary
computational overhead when mapping sparse multivectors using outermorphisms.

In this work, we propose an alternative approach for mapping arbitrary
multivectors called Online Basis-Mapping Method (OBMM). OBMM effectively
overcomes memory limitations while being reasonably efficient computationally
for high-dimensional GAs. The next section describes our approach
in full details.

\section{Proposed Approach}

\subsection{Online Basis-Mapping Method}

Algorithm \ref{alg:outermorphism-map} summarizes the proposed OBMM
procedure $Y\leftarrow\mathtt{OutermrphismMap}\left(\left\{ \boldsymbol{t}_{j}\right\} ,X\right)$
for efficiently computing outermorphisms of multivectors. The basic
idea behind OBMM is to online-compute the outermorphism $T_{i}=\overline{\mathbf{T}}\left[E_{i}\right]$
of basis blades $E_{i}$ while traversing the BTR of input multivector
$X$ to exploit its sparsity, if present. Inputs to the procedure
are the set of mapped basis vectors $\boldsymbol{t}_{k}=\overline{\mathbf{T}}\left[e_{k}\right]$
which fully defines a given outermorphism $\overline{\mathbf{T}}$,
and a given multivector $X$ to be mapped as $Y=\overline{\mathbf{T}}\left[X\right]$.
Input multivector $X$ is represented as a binary tree on the domain
GACF $\mathcal{E}\left(\boldsymbol{E}_{1}^{n},\mathbf{A}_{\mathcal{E}}\right)$
while output $Y$ is represented on the GACF $\mathcal{F}\left(\boldsymbol{F}_{1}^{m},\mathbf{A}_{\mathcal{F}}\right)$.

The procedure begins by initializing output multivector $Y$ to zero
and creating two stacks $S_{X}$ and $S_{T}$. Stack $S_{X}$ is used
to traverse the BTR of input multivector $X$, while stack $S_{T}$
is used for online computation and storage of $k$-vectors $T_{i}=\overline{\mathbf{T}}\left[E_{i}\right]$.
The main loop begins at step 6 until $S_{X}$ becomes empty when all
leaf nodes of $X$ are visited. Inside the main loop, the current
node $N_{X}$ and $k$-vector $T$ are popped from $S_{X}$ and $S_{T}$.
If $N_{X}$ is a leaf node, the output multivector $Y$ is updated
by adding $vT$ and the loop is continued, where $v$ is the scalar
coefficient value of the current leaf node $N_{X}$. At this stage
in the procedure, $T$ is the outer product of zero or more vectors
from the set $\left\{ \boldsymbol{t}_{j}\right\} $ as we will see
shortly from the following steps. If, on the other hand, node $N_{X}$
is an internal node, we push new values into stacks $S_{X}$ and $S_{T}$
accordingly. If $N_{X}$ has a 0-child node, we push the 0-child node
into $S_{X}$ and push $T$ without change into $S_{T}$. Finally,
if $N_{X}$ has a 1-child node, we push the 1-child node into $S_{X}$
and compute then push $\boldsymbol{t}_{k}\wedge T$ into $S_{T}$.

\begin{figure}
\noindent \begin{centering}
\includegraphics[width=7.5cm]{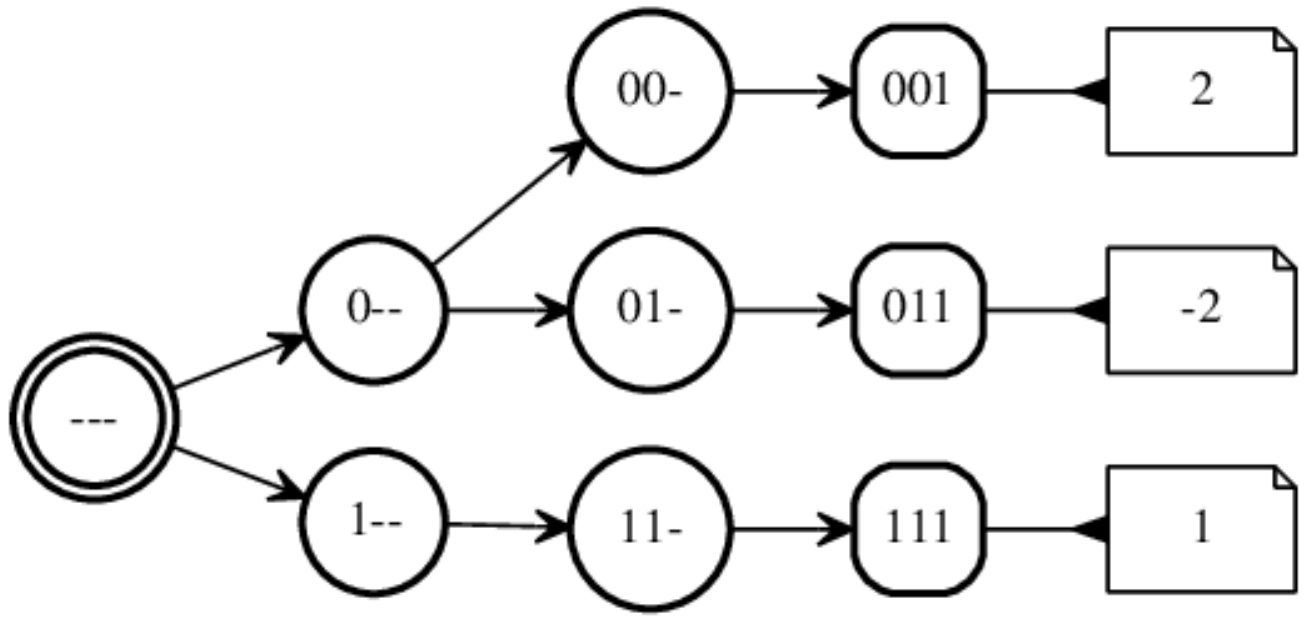}
\par\end{centering}
\caption{BTR of multivector $X=2e_{0}-2e_{01}+e_{012}$\label{fig:Example-Multivectors}}
\label{fig:tree-multivector-lcp-example}
\end{figure}

As an illustrative example, assume we have a 3-dimensional Euclidean
GACF $\mathcal{E}\left(\boldsymbol{E}_{1}^{n},\mathbf{A}_{\mathcal{E}}\right)$
with basis vectors $\left\langle e_{0},e_{1},e_{2}\right\rangle $.
Figure \ref{fig:Example-Multivectors} shows the binary tree representation
for an input multivector $X=2e_{0}-2e_{01}+e_{012}$. Figure \ref{fig:Execution-Listing-Example}
tracks the computational steps of $\overline{\mathbf{T}}\left[X\right]$
using Algorithm \ref{alg:outermorphism-map}.

\begin{algorithm}
\caption{$Y\leftarrow\mathtt{OutermrphismMap}\left(\left\{ \boldsymbol{t}_{j}\right\} ,X\right)$:
Computes the mapping $Y$ of a multivector $X$ under outermorphism
$\overline{\mathbf{T}}:\mathcal{E}\rightarrow\mathcal{F}$ defined
on GACFs $\mathcal{E}\left(\boldsymbol{E}_{1}^{n},\mathbf{A}_{\mathcal{E}}\right)$,
$\mathcal{F}\left(\boldsymbol{F}_{1}^{m},\mathbf{A}_{\mathcal{F}}\right)$
using mapping vectors $\boldsymbol{t}_{j}=\overline{\mathbf{T}}\left[e_{j}\right]$,
$j=0,1,\ldots,n-1$.}
\label{alg:outermorphism-map}
\begin{enumerate}
\item {\small{}Initialize output multivector $Y\leftarrow0$.}{\small\par}
\item {\small{}Initialize stack $S_{X}$ to traverse BTR nodes of input
multivector $X$.}{\small\par}
\item {\small{}Initialize stack $S_{T}$ to compute and store $k$-vectors
$T_{i}=\overline{\mathbf{T}}\left[E_{i}\right]$ of outermorphism
$\overline{\mathbf{T}}$.}{\small\par}
\item {\small{}Push root node of input multivector $X$ into stack $S_{X}$.}{\small\par}
\item {\small{}Push $0$-vector $T_{0}=1$ into stack $S_{T}$.}{\small\par}
\item {\small{}While stack $S_{X}$ is not empty, do steps 7-17:}{\small\par}
\item {\small{}\quad{}Pop top of stack $S_{X}$ into node $N_{X}$.}{\small\par}
\item {\small{}\quad{}Pop top of stack $S_{T}$ into $k$-vector $T$.}{\small\par}
\item {\small{}\quad{}If node $N_{X}$ is a leaf node do steps 10-11:}{\small\par}
\item {\small{}\quad{}\quad{}Update $Y\leftarrow Y+vT$ where $v=\mathtt{ScalarValue}\left(N_{X}\right)$.}{\small\par}
\item {\small{}\quad{}\quad{}Continue loop at step 6.}{\small\par}
\item {\small{}\quad{}If $N_{X}$ has a 0-child do steps 13-14:}{\small\par}
\item {\small{}\quad{}\quad{}Push $\mathtt{Child_{0}}\left(N_{X}\right)$
into stack $S_{X}$.}{\small\par}
\item {\small{}\quad{}\quad{}Push $T$ into stack $S_{T}$.}{\small\par}
\item {\small{}\quad{}If $N_{X}$ has a 1-child do steps 16-17:}{\small\par}
\item {\small{}\quad{}\quad{}Push $\mathtt{Child_{1}}\left(N_{X}\right)$
into stack $S_{X}$.}{\small\par}
\item {\small{}\quad{}\quad{}Push $\boldsymbol{t}_{j}\wedge T$ into stack
$S_{T}$, where $j=\mathtt{TreeDepth}\left(N_{X}\right)-1$.}{\small\par}
\item {\small{}Return final result in $Y$.}{\small\par}
\end{enumerate}
\end{algorithm}

\begin{figure}
\begin{singlespace}
\begin{raggedright}
\textbf{\small{}Initialize Stacks:}{\small\par}
\par\end{raggedright}
\begin{raggedright}
\textbf{\small{}\quad{}$\mathtt{Push}\left(X_{---},1\right)\,\mathtt{into}\,(S_{X},S_{T})$}{\small\par}
\par\end{raggedright}
\begin{raggedright}
\textbf{\small{}\quad{}}{\small\par}
\par\end{raggedright}
\begin{raggedright}
\textbf{\small{}Iteration 1:}{\small\par}
\par\end{raggedright}
\begin{raggedright}
\textbf{\small{}\quad{}$\mathtt{Pop}\left(X_{---},1\right)\,\mathtt{from}\,\left(S_{X},S_{T}\right)\,\mathtt{into}\,\left(N_{X},T\right)$}{\small\par}
\par\end{raggedright}
\begin{raggedright}
\textbf{\small{}\quad{}Internal node; update stacks $S_{X},S_{T}$:}{\small\par}
\par\end{raggedright}
\begin{raggedright}
\textbf{\small{}\quad{}\quad{}$\mathtt{Push}\left(X_{0--},1\right)\,\mathtt{into}\,\left(S_{X},S_{T}\right)$}{\small\par}
\par\end{raggedright}
\begin{raggedright}
\textbf{\small{}\quad{}\quad{}$\mathtt{Push}\left(X_{1--},\boldsymbol{t}_{2}\right)\,\mathtt{into}\,\left(S_{X},S_{T}\right)$}{\small\par}
\par\end{raggedright}
\begin{raggedright}
\textbf{\small{}\quad{}}{\small\par}
\par\end{raggedright}
\begin{raggedright}
\textbf{\small{}Iteration 2:}{\small\par}
\par\end{raggedright}
\begin{raggedright}
\textbf{\small{}\quad{}$\mathtt{Pop}\left(X_{1--},\boldsymbol{t}_{2}\right)\,\mathtt{from}\,\left(S_{X},S_{T}\right)\,\mathtt{into}\,\left(N_{X},T\right)$}{\small\par}
\par\end{raggedright}
\begin{raggedright}
\textbf{\small{}\quad{}Internal nodes; update stacks $S_{X},S_{T}$:}{\small\par}
\par\end{raggedright}
\begin{raggedright}
\textbf{\small{}\quad{}\quad{}$\mathtt{Push}\left(X_{11-},\boldsymbol{t}_{1}\wedge\boldsymbol{t}_{2}\right)\,\mathtt{into}\,\left(S_{X},S_{T}\right)$}{\small\par}
\par\end{raggedright}
\begin{raggedright}
\textbf{\small{}\quad{}}{\small\par}
\par\end{raggedright}
\begin{raggedright}
\textbf{\small{}Iteration 3:}{\small\par}
\par\end{raggedright}
\begin{raggedright}
\textbf{\small{}\quad{}$\mathtt{Pop}\left(X_{11-},\boldsymbol{t}_{1}\wedge\boldsymbol{t}_{2}\right)\,\mathtt{from}\,\left(S_{X},S_{T}\right)\,\mathtt{into}\,\left(N_{X},T\right)$}{\small\par}
\par\end{raggedright}
\begin{raggedright}
\textbf{\small{}\quad{}Internal node; update stacks $S_{X},S_{T}$:}{\small\par}
\par\end{raggedright}
\begin{raggedright}
\textbf{\small{}\quad{}\quad{}$\mathtt{Push}\left(X_{111},\boldsymbol{t}_{0}\wedge\boldsymbol{t}_{1}\wedge\boldsymbol{t}_{2}\right)\,\mathtt{into}\,\left(S_{X},S_{T}\right)$}{\small\par}
\par\end{raggedright}
\begin{raggedright}
\textbf{\small{}\quad{}}{\small\par}
\par\end{raggedright}
\begin{raggedright}
\textbf{\small{}Iteration 4:}{\small\par}
\par\end{raggedright}
\begin{raggedright}
\textbf{\small{}\quad{}$\mathtt{Pop}\left(X_{111},\boldsymbol{t}_{0}\wedge\boldsymbol{t}_{1}\wedge\boldsymbol{t}_{2}\right)\,\mathtt{from}\,\left(S_{X},S_{T}\right)\,\mathtt{into}\,\left(N_{X},T\right)$}{\small\par}
\par\end{raggedright}
\begin{raggedright}
\textbf{\small{}\quad{}Leaf node; update output multivector $Y\leftarrow Y+\left(1\right)\boldsymbol{t}_{0}\wedge\boldsymbol{t}_{1}\wedge\boldsymbol{t}_{2}$}{\small\par}
\par\end{raggedright}
\begin{raggedright}
\textbf{\small{}\quad{}}{\small\par}
\par\end{raggedright}
\begin{raggedright}
\textbf{\small{}Iteration 5:}{\small\par}
\par\end{raggedright}
\begin{raggedright}
\textbf{\small{}\quad{}$\mathtt{Pop}\left(X_{0--},1\right)\,\mathtt{from}\,\left(S_{X},S_{T}\right)\,\mathtt{into}\,\left(N_{X},T\right)$}{\small\par}
\par\end{raggedright}
\begin{raggedright}
\textbf{\small{}\quad{}Internal node; update stacks $S_{X},S_{T}$:}{\small\par}
\par\end{raggedright}
\begin{raggedright}
\textbf{\small{}\quad{}\quad{}$\mathtt{Push}\left(X_{00-},1\right)\,\mathtt{into}\,\left(S_{X},S_{T}\right)$}{\small\par}
\par\end{raggedright}
\begin{raggedright}
\textbf{\small{}\quad{}\quad{}$\mathtt{Push}\left(X_{01-},\boldsymbol{t}_{1}\right)\,\mathtt{into}\,\left(S_{X},S_{T}\right)$}{\small\par}
\par\end{raggedright}
\begin{raggedright}
\textbf{\small{}\quad{}}{\small\par}
\par\end{raggedright}
\begin{raggedright}
\textbf{\small{}Iteration 6:}{\small\par}
\par\end{raggedright}
\begin{raggedright}
\textbf{\small{}\quad{}$\mathtt{Pop}\left(X_{01-},\boldsymbol{t}_{1}\right)\,\mathtt{from}\,\left(S_{X},S_{T}\right)\,\mathtt{into}\,\left(N_{X},T\right)$}{\small\par}
\par\end{raggedright}
\begin{raggedright}
\textbf{\small{}\quad{}Internal node; update stacks $S_{X},S_{T}$:}{\small\par}
\par\end{raggedright}
\begin{raggedright}
\textbf{\small{}\quad{}\quad{}$\mathtt{Push}\left(X_{011},\boldsymbol{t}_{0}\wedge\boldsymbol{t}_{1}\right)\,\mathtt{into}\,\left(S_{X},S_{T}\right)$}{\small\par}
\par\end{raggedright}
\begin{raggedright}
\textbf{\small{}\quad{}}{\small\par}
\par\end{raggedright}
\begin{raggedright}
\textbf{\small{}Iteration 7:}{\small\par}
\par\end{raggedright}
\begin{raggedright}
\textbf{\small{}\quad{}$\mathtt{Pop}\left(X_{011},\boldsymbol{t}_{0}\wedge\boldsymbol{t}_{1}\right)\,\mathtt{from}\,\left(S_{X},S_{T}\right)\,\mathtt{into}\,\left(N_{X},T\right)$}{\small\par}
\par\end{raggedright}
\begin{raggedright}
\textbf{\small{}\quad{}Leaf node; update output multivector $Y\leftarrow Y+\left(-2\right)\boldsymbol{t}_{0}\wedge\boldsymbol{t}_{1}$}{\small\par}
\par\end{raggedright}
\begin{raggedright}
\textbf{\small{}\quad{}}{\small\par}
\par\end{raggedright}
\begin{raggedright}
\textbf{\small{}Iteration 8:}{\small\par}
\par\end{raggedright}
\begin{raggedright}
\textbf{\small{}\quad{}$\mathtt{Pop}\left(X_{00-},1\right)\,\mathtt{from}\,\left(S_{X},S_{T}\right)\,\mathtt{into}\,\left(N_{X},T\right)$}{\small\par}
\par\end{raggedright}
\begin{raggedright}
\textbf{\small{}\quad{}Internal node; update stacks $S_{X},S_{T}$:}{\small\par}
\par\end{raggedright}
\begin{raggedright}
\textbf{\small{}\quad{}\quad{}$\mathtt{Push}\left(X_{001},\boldsymbol{t}_{0}\right)\,\mathtt{into}\,\left(S_{X},S_{T}\right)$}{\small\par}
\par\end{raggedright}
\begin{raggedright}
\textbf{\small{}\quad{}}{\small\par}
\par\end{raggedright}
\begin{raggedright}
\textbf{\small{}Iteration 9:}{\small\par}
\par\end{raggedright}
\begin{raggedright}
\textbf{\small{}\quad{}$\mathtt{Pop}\left(X_{001},\boldsymbol{t}_{0}\right)\,\mathtt{from}\,\left(S_{X},S_{T}\right)\,\mathtt{into}\,\left(N_{X},T\right)$}{\small\par}
\par\end{raggedright}
\begin{raggedright}
\textbf{\small{}\quad{}Leaf node; update output multivector $Y\leftarrow Y+\left(2\right)\boldsymbol{t}_{0}$}{\small\par}
\par\end{raggedright}
\end{singlespace}
\caption{Listing of iterations of outermorphism computation $\overline{\mathbf{T}}\left[X\right]$\label{fig:Execution-Listing-Example}}
\end{figure}

\subsection{Efficient Implementation Details}

The computational bottleneck of Algorithm \ref{alg:outermorphism-map}
is at step 17 when computing the outer product $\boldsymbol{t}_{k}\wedge T$
between vector $\boldsymbol{t}_{k}$ and $k$-vector $T$. Implementing
this outer product using common methods is not feasible. If we use
a lookup table for the outer product we would need too much memory
for higher-dimensional GAs. On the other hand, if we use simple loops
or tree-based procedures \cite{Breuils2017,Breuils2017a,breuils:tel-02085820}
for computing the outer product performance would suffer significantly.
In this work, we implemented the outer product computation $\boldsymbol{t}_{k}\wedge T$
using simple code generation. We first created a special class for
holding $k$-vector coefficients as shown in Figure \ref{fig:Code-Listing-1}
and then generated highly efficient functions for computing the outer
product. Figure \ref{fig:Code-Listing-2} lists parts of the main
function which selects how to compute the outer product depending
on the GA dimension and the grade of $k$-vector $T$. For each GA
dimension $n$, a set of $n-1$ computational functions were generated
to efficiently compute the desired outer product. Each computational
function is specialized in a specific GA dimension and $k$-vector
grade within the GA. Figure \ref{fig:Code-Listing-3} lists the two
generated functions for 3-dimensional GAs. Such approach requires
no additional memory, aside from inputs and output, for $\boldsymbol{t}_{k}\wedge T$
outer product computations, while significantly reducing computational
time compared to other approaches.
\begin{figure}
\begin{singlespace}
\begin{raggedright}
\texttt{\small{}public class GaNumKVector \{}{\small\par}
\par\end{raggedright}
\begin{raggedright}
\texttt{\small{}\quad{}public double{[}{]} ScalarValuesArray \{ get;
\}}{\small\par}
\par\end{raggedright}
\begin{raggedright}
\texttt{\small{}\quad{}public int Grade \{ get; \}}{\small\par}
\par\end{raggedright}
\begin{raggedright}
\texttt{\small{}\quad{}public int VSpaceDimension \{ get; \}}{\small\par}
\par\end{raggedright}
\begin{raggedright}
\texttt{\small{}\quad{}}{\small\par}
\par\end{raggedright}
\begin{raggedright}
\texttt{\small{}\quad{}public GaNumKVector(int vSpaceDim, int grade,
double{[}{]} scalarValuesArray) \{}{\small\par}
\par\end{raggedright}
\begin{raggedright}
\texttt{\small{}\quad{}\quad{}VSpaceDimension = vSpaceDim;}{\small\par}
\par\end{raggedright}
\begin{raggedright}
\texttt{\small{}\quad{}\quad{}Grade = grade;}{\small\par}
\par\end{raggedright}
\begin{raggedright}
\texttt{\small{}\quad{}\quad{}ScalarValuesArray = scalarValuesArray;}{\small\par}
\par\end{raggedright}
\begin{raggedright}
\texttt{\small{}\quad{}\}}{\small\par}
\par\end{raggedright}
\begin{raggedright}
\texttt{\small{}\}}{\small\par}
\par\end{raggedright}
\end{singlespace}
\caption{Simplified class definition for holding $k$-vector information\label{fig:Code-Listing-1}}
\end{figure}
\begin{figure}
\begin{singlespace}
\begin{raggedright}
\texttt{\small{}public GaNumKVector VectorKVectorOp(GaNumKVector vector,
GaNumKVector kVector) \{}{\small\par}
\par\end{raggedright}
\begin{raggedright}
\quad{}\texttt{\small{}var vSpaceDim = kVector.VSpaceDimension;}{\small\par}
\par\end{raggedright}
\begin{raggedright}
\quad{}\texttt{\small{}var grade = kVector.Grade;}{\small\par}
\par\end{raggedright}
\begin{raggedright}
\quad{}
\par\end{raggedright}
\begin{raggedright}
\quad{}\texttt{\small{}//Compute the outer product in a 2-dimensional
GA}{\small\par}
\par\end{raggedright}
\begin{raggedright}
\quad{}\texttt{\small{}if (vSpaceDim == 2)} \texttt{\small{}\{}{\small\par}
\par\end{raggedright}
\begin{raggedright}
\quad{}\quad{}\texttt{\small{}if (grade == 0) return vector;}{\small\par}
\par\end{raggedright}
\begin{raggedright}
\quad{}\quad{}\texttt{\small{}if (grade == 1) return VectorKVectorOp\_2\_1(vector,
kVector);}{\small\par}
\par\end{raggedright}
\begin{raggedright}
\quad{}\texttt{\small{}\}}{\small\par}
\par\end{raggedright}
\begin{raggedright}
\quad{}
\par\end{raggedright}
\begin{raggedright}
\quad{}\texttt{\small{}//Compute the outer product in a 3-dimensional
GA}{\small\par}
\par\end{raggedright}
\begin{raggedright}
\quad{}\texttt{\small{}if (vSpaceDim == 3)} \texttt{\small{}\{}{\small\par}
\par\end{raggedright}
\begin{raggedright}
\quad{}\quad{}\texttt{\small{}if (grade == 0) return vector;}{\small\par}
\par\end{raggedright}
\begin{raggedright}
\quad{}\quad{}\texttt{\small{}if (grade == 1) return VectorKVectorOp\_3\_1(vector,
kVector);}{\small\par}
\par\end{raggedright}
\begin{raggedright}
\quad{}\quad{}\texttt{\small{}if (grade == 2) return VectorKVectorOp\_3\_2(vector,
kVector);}{\small\par}
\par\end{raggedright}
\begin{raggedright}
\quad{}\texttt{\small{}\}}{\small\par}
\par\end{raggedright}
\begin{raggedright}
\quad{}
\par\end{raggedright}
\begin{raggedright}
\quad{}\texttt{\small{}//Compute the outer product in a 4-dimensional
GA}{\small\par}
\par\end{raggedright}
\begin{raggedright}
\quad{}\texttt{\small{}if (vSpaceDim == 4) \{}{\small\par}
\par\end{raggedright}
\begin{raggedright}
\quad{}\quad{}\texttt{\small{}if (grade == 0) return vector;}{\small\par}
\par\end{raggedright}
\begin{raggedright}
\quad{}\quad{}\texttt{\small{}if (grade == 1) return VectorKVectorOp\_4\_1(vector,
kVector);}{\small\par}
\par\end{raggedright}
\begin{raggedright}
\quad{}\quad{}\texttt{\small{}if (grade == 2) return VectorKVectorOp\_4\_2(vector,
kVector);}{\small\par}
\par\end{raggedright}
\begin{raggedright}
\quad{}\quad{}\texttt{\small{}if (grade == 3) return VectorKVectorOp\_4\_3(vector,
kVector);}{\small\par}
\par\end{raggedright}
\begin{raggedright}
\quad{}\texttt{\small{}\}}{\small\par}
\par\end{raggedright}
\begin{raggedright}
\quad{}\texttt{\small{}.}{\small\par}
\par\end{raggedright}
\begin{raggedright}
\quad{}\texttt{\small{}.}{\small\par}
\par\end{raggedright}
\begin{raggedright}
\quad{}\texttt{\small{}.}{\small\par}
\par\end{raggedright}
\begin{raggedright}
\texttt{\small{}\}}{\small\par}
\par\end{raggedright}
\end{singlespace}
\caption{Main subroutine definition for selecting a specific function to compute
the outer product of vector $\boldsymbol{t}_{k}$ with $k$-vector
$T$\label{fig:Code-Listing-2}}
\end{figure}
\begin{figure}
\begin{singlespace}
\begin{raggedright}
\texttt{\small{}//Compute the outer product of a vector and a 1-vector}{\small\par}
\par\end{raggedright}
\begin{raggedright}
\texttt{\small{}//in a 3-dimensional GA; the result is always a 2-vector}{\small\par}
\par\end{raggedright}
\begin{raggedright}
\texttt{\small{}private GaNumKVector VectorKVectorOp\_3\_1(GaNumKVector
vector, GaNumKVector kVector) \{}{\small\par}
\par\end{raggedright}
\begin{raggedright}
\quad{}\texttt{\small{}var vectorArray = vector.ScalarValuesArray;}{\small\par}
\par\end{raggedright}
\begin{raggedright}
\quad{}\texttt{\small{}var kVectorArray = kVector.ScalarValuesArray;}{\small\par}
\par\end{raggedright}
\begin{raggedright}
\quad{}\texttt{\small{}var resultArray = new double{[}3{]};}{\small\par}
\par\end{raggedright}
\begin{raggedright}
\quad{}
\par\end{raggedright}
\begin{raggedright}
\quad{}\texttt{\small{}var value = 0.0d;}{\small\par}
\par\end{raggedright}
\begin{raggedright}
\quad{}\texttt{\small{}value += vectorArray{[}0{]} {*} kVectorArray{[}1{]};}{\small\par}
\par\end{raggedright}
\begin{raggedright}
\quad{}\texttt{\small{}value -= vectorArray{[}1{]} {*} kVectorArray{[}0{]};}{\small\par}
\par\end{raggedright}
\begin{raggedright}
\quad{}\texttt{\small{}resultArray{[}0{]} = value;}{\small\par}
\par\end{raggedright}
\begin{raggedright}
\quad{}
\par\end{raggedright}
\begin{raggedright}
\quad{}\texttt{\small{}value = 0.0d;}{\small\par}
\par\end{raggedright}
\begin{raggedright}
\quad{}\texttt{\small{}value += vectorArray{[}0{]} {*} kVectorArray{[}2{]};}{\small\par}
\par\end{raggedright}
\begin{raggedright}
\quad{}\texttt{\small{}value -= vectorArray{[}2{]} {*} kVectorArray{[}0{]};}{\small\par}
\par\end{raggedright}
\begin{raggedright}
\quad{}\texttt{\small{}resultArray{[}1{]} = value;}{\small\par}
\par\end{raggedright}
\begin{raggedright}
\quad{}
\par\end{raggedright}
\begin{raggedright}
\quad{}\texttt{\small{}value = 0.0d;}{\small\par}
\par\end{raggedright}
\begin{raggedright}
\quad{}\texttt{\small{}value += vectorArray{[}1{]} {*} kVectorArray{[}2{]};}{\small\par}
\par\end{raggedright}
\begin{raggedright}
\quad{}\texttt{\small{}value -= vectorArray{[}2{]} {*} kVectorArray{[}1{]};}{\small\par}
\par\end{raggedright}
\begin{raggedright}
\quad{}\texttt{\small{}resultArray{[}2{]} = value;}{\small\par}
\par\end{raggedright}
\begin{raggedright}
\quad{}
\par\end{raggedright}
\begin{raggedright}
\quad{}\texttt{\small{}return new GaNumKVector(3, 2, resultArray);}{\small\par}
\par\end{raggedright}
\begin{raggedright}
\texttt{\small{}\}}{\small\par}
\par\end{raggedright}
\begin{raggedright}
\quad{}
\par\end{raggedright}
\begin{raggedright}
\texttt{\small{}//Compute the outer product of a vector and a 2-vector}{\small\par}
\par\end{raggedright}
\begin{raggedright}
\texttt{\small{}//in a 3-dimensional GA; the result is always a 3-vector}{\small\par}
\par\end{raggedright}
\begin{raggedright}
\texttt{\small{}private GaNumKVector VectorKVectorOp\_3\_2(GaNumKVector
vector, GaNumKVector kVector) \{}{\small\par}
\par\end{raggedright}
\begin{raggedright}
\quad{}\texttt{\small{}var vectorArray = vector.ScalarValuesArray;}{\small\par}
\par\end{raggedright}
\begin{raggedright}
\quad{}\texttt{\small{}var kVectorArray = kVector.ScalarValuesArray;}{\small\par}
\par\end{raggedright}
\begin{raggedright}
\quad{}\texttt{\small{}var resultArray = new double{[}1{]};}{\small\par}
\par\end{raggedright}
\begin{raggedright}
\quad{}
\par\end{raggedright}
\begin{raggedright}
\quad{}\texttt{\small{}var value = 0.0d;}{\small\par}
\par\end{raggedright}
\begin{raggedright}
\quad{}\texttt{\small{}value += vectorArray{[}0{]} {*} kVectorArray{[}2{]};}{\small\par}
\par\end{raggedright}
\begin{raggedright}
\quad{}\texttt{\small{}value -= vectorArray{[}1{]} {*} kVectorArray{[}1{]};}{\small\par}
\par\end{raggedright}
\begin{raggedright}
\quad{}\texttt{\small{}value += vectorArray{[}2{]} {*} kVectorArray{[}0{]};}{\small\par}
\par\end{raggedright}
\begin{raggedright}
\quad{}\texttt{\small{}resultArray{[}0{]} = value;}{\small\par}
\par\end{raggedright}
\begin{raggedright}
\quad{}
\par\end{raggedright}
\begin{raggedright}
\quad{}\texttt{\small{}return new GaNumKVector(3, 3, resultArray);}{\small\par}
\par\end{raggedright}
\begin{raggedright}
\texttt{\small{}\}}{\small\par}
\par\end{raggedright}
\end{singlespace}
\caption{Function definitions for computing the outer product of vector $\boldsymbol{t}_{k}$
with $k$-vector $T$ in any 3-dimensional GA\label{fig:Code-Listing-3}}
\end{figure}

\section{Results and Discussion}

\begin{figure}
\begin{centering}
\includegraphics[width=7.5cm]{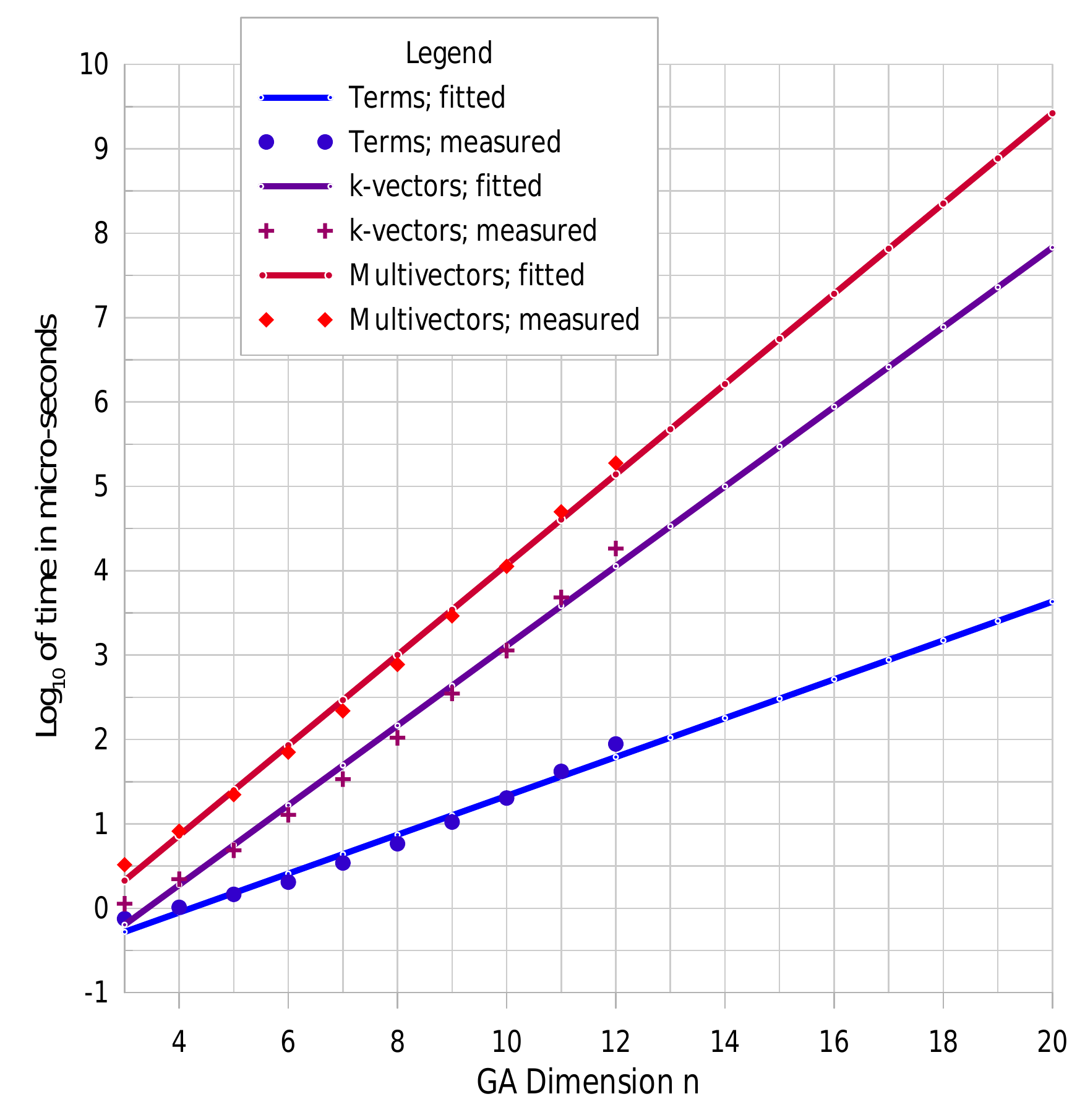}
\par\end{centering}
\caption{\label{fig:multivector-sparsity-effect}Effect of multivector sparsity
on computation time of outermorphisms using OBMM}
\end{figure}

We created two implementations to test the usefulness of the proposed
approach. The first implementation, based on CBMM, is by pre-computing
all $k$-vectors $T_{i}$ and storing them into a simple array indexed
by $i$. The outermorphism of a multivector $X=\sum_{i=0}^{2^{n}-1}x_{i}E_{i}$
is then computed using a simple loop $\overline{\mathbf{T}}\left[X\right]=\sum_{i=0,x_{i}\neq0}^{2^{n}-1}x_{i}T_{i}$
where the loop is over non-zero terms in $X$ having $x_{i}\neq0$.
It is important to note that CBMM doesn't require the use of BTR for
multivectors. The second implementation is the proposed OBMM procedure
combined with code generation as described previously. Experimental
trials to measure time performance were made for GACFs with dimension
$n$ ranging from 3 to 12 on an i7-class machine with 8 GBytes of
memory. Three kinds of multivectors were used to show the effect of
multivector sparsity on computation time. The first kind contains
full multivectors with no missing terms, which is the least sparse
kind of multivectors. The second kind contains $k$-vectors of all
grades $0\leq k\leq n$. The third kind contains single-term multivectors,
which is the most sparse kind. Table \ref{tab:average-time} summarizes
average computation time, in micro-seconds, for all trials. Data in
this table suggests an exponential complexity growth in time as a
function of GA dimension $n$. We applied simple exponential curve-fitting
to find the parameters of an exponential function $cb^{n}$ that approximates
each column of data in the table. Constant $c$ and base $b$ parameters
for each approximating function are shown at the bottom of the corresponding
column. The effect of multivector sparsity on computation time is
illustrated in Figure \ref{fig:multivector-sparsity-effect}. For
computing outermorphisms of multivectors, the base parameter $b$
decreases considerably with increased multivector sparsity as shown
in the figure. As most important computations in GA involve blades,
the case of $k$-vectors should be taken as the dominant one when
designing and optimizing GA computations.

Another set of trials were made to measure memory requirements for
defining and computing outermorphisms for GACFs with dimension $n$
ranging from 3 to 15. Table \ref{tab:memory} shows the results of
this set of trials. The column labeled 'CBMM Definition' lists the
total memory required for defining an outermorphism using CBMM for
various values of GA dimension $n$. Memory requirements for CBMM
is proportional to the function $\sum_{k=0}^{n}\left(\begin{array}{c}
n\\
k
\end{array}\right)^{2}$ equal to the number of scalars stored for all $T_{i}$. On the other
hand, memory needed for the definition of OBMM, as shown in the following
column 'OBMM Definition', is proportional to $n^{2}$; thus giving
many orders of magnitude lower memory requirements compared to CBMM.
When mapping a multivector using CBMM, no additional memory is needed.
However, for the OBMM approach, additional memory is required for
stack $S_{T}$ holding online computations of $\boldsymbol{t}_{k}\wedge T$
as explained earlier. The 'OBMM Mapping' column of Table \ref{tab:memory}
lists the maximum memory required for stack $S_{T}$. The maximum
memory listed in this column is only needed when performing multivector
mappings, and is freed afterwards. The memory listed in the CBMM and
OBMM definition columns, on the other hand, are needed for the full
life time of the outermorphism. The last 3 columns in the table display
memory required for storing BTRs of multivectors. Because BTRs are
not needed for the CBMM approach, they are considered additional memory
overhead needed for OBMM. Nevertheless, combined memory requirements
of OBMM definition, mapping, and multivector BTR are significantly
small compared to CBMM memory requirements.

From the measured time data and approximation function parameters
in Table \ref{tab:average-time}, it is clear that time performances
of both OBMM and CBMM are very close. OBMM has the additional benefit
of requiring significantly smaller memory storage compared to the
huge quantity of memory needed for CBMM as seen from Table \ref{tab:memory}.
For GAs with dimension $n>14$, memory requirements of CBMM are limiting,
while the difference in performance with OBMM is practically negligible.
On the other hand, data required to fully define an outermorphism
operator in OBMM is a small square matrix of size $n\times n$, only
needing $kn^{2}$ bytes in memory, where $k\geq8$ depends on implementation
specifics. OBMM can thus be used to compute related outermorphisms
based on their $n\times n$ basis vector mapping matrices alone. For
example, given an outermorphism defined by its basis vector mapping
matrix, we can use the matrix to compute its inverse outermorphism,
find its adjoint, factorize it into related outermorphisms, or compose
several outermorphisms into a single one. All such standard matrix
computations are performed on the much smaller $n\times n$ matrices
fully defining the outermorphisms without any need to store or manipulate
the exponentially larger $2^{n}\times2^{n}$ multivector linear mapping
matrices. This opens the door for using standard linear algebra libraries
for efficiently defining, analyzing, and relating outermorphisms;
thus giving a much wider field for computing with outermorphisms in
practice.

\begin{table}
\centering{}\caption{Average time required for computing the outermorphism of multivectors
in micro-seconds using OBMM and CBMM for various GA dimensions and
multivector sparsity, and its exponential curve fitting approximation
parameters.\label{tab:average-time}}
\begin{tabular}{|c|r|r|r|r|r|r|}
\hline 
\multirow{2}{*}{\textbf{\small{}$n$}} & \multicolumn{2}{c|}{\textbf{\small{}Multivectors}} & \multicolumn{2}{c|}{{\small{}$k$}\textbf{\small{}-vectors}} & \multicolumn{2}{c|}{\textbf{\small{}Terms}}\tabularnewline
\cline{2-7} \cline{3-7} \cline{4-7} \cline{5-7} \cline{6-7} \cline{7-7} 
 & \textbf{\small{}OBMM} & \textbf{\small{}CBMM} & \textbf{\small{}OBMM} & \textbf{\small{}CBMM} & \textbf{\small{}OBMM} & \textbf{\small{}CBMM}\tabularnewline
\hline 
{\small{}3} & {\small{}3.27} & {\small{}2.73} & {\small{}1.13} & {\small{}0.77} & {\small{}0.75} & {\small{}0.50}\tabularnewline
\hline 
{\small{}4} & {\small{}8.20} & {\small{}6.24} & {\small{}2.21} & {\small{}1.49} & {\small{}1.03} & {\small{}0.69}\tabularnewline
\hline 
{\small{}5} & {\small{}22.19} & {\small{}18.21} & {\small{}4.86} & {\small{}3.34} & {\small{}1.46} & {\small{}1.02}\tabularnewline
\hline 
{\small{}6} & {\small{}70.71} & {\small{}54.49} & {\small{}12.81} & {\small{}9.14} & {\small{}2.04} & {\small{}1.53}\tabularnewline
\hline 
{\small{}7} & {\small{}218.22} & {\small{}196.62} & {\small{}33.88} & {\small{}27.64} & {\small{}3.44} & {\small{}2.38}\tabularnewline
\hline 
{\small{}8} & {\small{}774.37} & {\small{}708.17} & {\small{}104.76} & {\small{}85.56} & {\small{}5.81} & {\small{}4.37}\tabularnewline
\hline 
{\small{}9} & {\small{}2,913.22} & {\small{}2,559.38} & {\small{}351.09} & {\small{}287.64} & {\small{}10.51} & {\small{}7.91}\tabularnewline
\hline 
{\small{}10} & {\small{}11,271.69} & {\small{}9,755.97} & {\small{}1,132.66} & {\small{}947.05} & {\small{}20.23} & {\small{}15.40}\tabularnewline
\hline 
{\small{}11} & {\small{}50,006.43} & {\small{}36,966.67} & {\small{}4,825.43} & {\small{}3,305.80} & {\small{}41.87} & {\small{}28.11}\tabularnewline
\hline 
{\small{}12} & {\small{}188,891.63} & {\small{}141,657.39} & {\small{}18,347.80} & {\small{}12,049.23} & {\small{}88.47} & {\small{}52.96}\tabularnewline
\hline 
\hline 
\multicolumn{7}{|c|}{\textbf{\small{}Exponential Curve Fitting Approximation $cb^{n}$}}\tabularnewline
\hline 
{\small{}$b$} & {\small{}3.4276} & {\small{}3.4168} & {\small{}2.9661} & {\small{}2.9864} & {\small{}1.6992} & {\small{}1.6956}\tabularnewline
\hline 
{\small{}$c$} & {\small{}0.0527} & {\small{}0.0443} & {\small{}0.0244} & {\small{}0.0171} & {\small{}0.1069} & {\small{}0.0759}\tabularnewline
\hline 
\end{tabular}
\end{table}

\begin{table}
\centering{}\caption{Memory required for defining and computing the outermorphism of multivectors
in bytes using OBMM and CBMM for various GA dimensions.\label{tab:memory}}
\begin{turn}{90}
\begin{tabular}{|c|c|c|c|c|c|c|}
\hline 
\multirow{2}{*}{\textbf{\small{}$n$}} & \textbf{\small{}CBMM} & \multicolumn{2}{c|}{\textbf{\small{}OBMM}} & \multicolumn{3}{c|}{\textbf{\small{}BTR}}\tabularnewline
\cline{3-7} \cline{4-7} \cline{5-7} \cline{6-7} \cline{7-7} 
 & \textbf{\small{}Definition} & \textbf{\small{}Definition} & \textbf{\small{}Mapping} & \textbf{\small{}Multivectors} & \textbf{\small{}k-vectors} & \textbf{\small{}Terms}\tabularnewline
\hline 
{\small{}3} & {\small{}416} & {\small{}336} & {\small{}224} & {\small{}380} & {\small{}232} & {\small{}120}\tabularnewline
\hline 
{\small{}4} & {\small{}1,012} & {\small{}472} & {\small{}336} & {\small{}768} & {\small{}452} & {\small{}148}\tabularnewline
\hline 
{\small{}5} & {\small{}2,856} & {\small{}640} & {\small{}540} & {\small{}1,540} & {\small{}808} & {\small{}176}\tabularnewline
\hline 
{\small{}6} & {\small{}9,004} & {\small{}840} & {\small{}940} & {\small{}3,080} & {\small{}1,572} & {\small{}204}\tabularnewline
\hline 
{\small{}7} & {\small{}30,608} & {\small{}1,072} & {\small{}1,768} & {\small{}6,156} & {\small{}2,872} & {\small{}232}\tabularnewline
\hline 
{\small{}8} & {\small{}109,188} & {\small{}1,336} & {\small{}3,480} & {\small{}12,304} & {\small{}5,620} & {\small{}260}\tabularnewline
\hline 
{\small{}9} & {\small{}401,336} & {\small{}1,632} & {\small{}6,972} & {\small{}24,596} & {\small{}10,440} & {\small{}288}\tabularnewline
\hline 
{\small{}10} & {\small{}1,502,716} & {\small{}1,960} & {\small{}14,028} & {\small{}49,176} & {\small{}20,524} & {\small{}316}\tabularnewline
\hline 
{\small{}11} & {\small{}5,692,704} & {\small{}2,320} & {\small{}28,784} & {\small{}98,332} & {\small{}38,576} & {\small{}344}\tabularnewline
\hline 
{\small{}12} & {\small{}21,731,652} & {\small{}2,712} & {\small{}58,352} & {\small{}196,640} & {\small{}76,068} & {\small{}372}\tabularnewline
\hline 
{\small{}13} & {\small{}83,401,512} & {\small{}3,136} & {\small{}120,412} & {\small{}393,252} & {\small{}144,184} & {\small{}400}\tabularnewline
\hline 
{\small{}14} & {\small{}321,326,124} & {\small{}3,592} & {\small{}243,964} & {\small{}786,472} & {\small{}284,900} & {\small{}428}\tabularnewline
\hline 
{\small{}15} & {\small{}1,241,726,704} & {\small{}4,080} & {\small{}503,672} & {\small{}1,572,908} & {\small{}543,448} & {\small{}456}\tabularnewline
\hline 
\hline 
\multicolumn{7}{|c|}{\textbf{\small{}Asymbtotically Approximating Functions}}\tabularnewline
\hline 
 & {\small{}$8\sum_{k=0}^{n}\left(\begin{array}{c}
n\\
k
\end{array}\right)^{2}$} & {\small{}$18n^{2}$} & {\small{}$21\left(1.936\right)^{n}$} & {\small{}$48\left(2\right)^{n}$} & {\small{}$32.3764\left(1.9087\right)^{n}$} & {\small{}$102.34\left(1.1124\right)^{n}$}\tabularnewline
\hline 
\end{tabular}
\end{turn}
\end{table}

\section{Conclusion}

In this work, we have presented a time-efficient, low-memory approach
for implementing outermorphisms. The basic idea behind the approach
is to online-compute the outermorphism $T_{i}=\overline{\mathbf{T}}\left[E_{i}\right]$
of basis blades $E_{i}$ while traversing a binary tree representation
of input multivector $X=\sum_{i=0}^{2^{n}-1}x_{i}E_{i}$ to effectively
exploit its sparsity, if present. We utilized simple code generation
to accelerate the processing of the computational bottleneck when
computing $k$-vectors $T_{i}$. Compared to typical approaches, which
pre-compute and store all $k$-vectors $T_{i}$, our approach requires
orders of magnitude less memory and performs comparably well regarding
computation time.

As a next step, we plan to make further acceleration by utilizing
CPU multi-threading or GPU parallel processing techniques when traversing
BTRs of high-dimension multivectors. In addition, we are currently
studying and developing similar techniques for efficiently computing
common bilinear products on multivectors. Combining efficient outermorphism
mapping with efficient products on multivectors is especially useful
for multivector computations on non-orthogonal GA coordinate frames.

\end{document}